\documentclass[12pt]{article}
\usepackage{amsmath}
\usepackage{amssymb}
\usepackage{array}
\usepackage{geometry}
\usepackage{graphicx}
\usepackage{enumerate}
\numberwithin{equation}{section}
\newtheorem{thm}{Theorem}[section]

\newtheorem{cor}{Corollary}[section]

\begin{document}
\numberwithin{equation}{section}
\begin{center}
{\bf \large  ON EXTENDED THERMONUCLEAR FUNCTIONS
 THROUGH PATHWAY MODEL}\\
\bigskip
{\bf Dilip Kumar}\\
{\small
Centre for Mathematical Sciences Pala Campus, Arunapuram P.O.,\\ 
Palai, Kerala  686 574, India\\
Email: dilipkumar.cms@gmail.com}\\
\bigskip
{\bf Hans J. Haubold}\\
{\small
Office for Outer Space Affairs, United Nations, Vienna International Centre\\
A-1400 Vienna, Austria\\
Email: hans.haubold@unvienna.org}
\end{center}
{ \small{\bf Abstract.}  The major problem in the cosmological
nucleosynthesis is the evaluation of the reaction rate.  The present
scenario is that the standard thermonuclear function in the
Maxwell-Boltzmann form is evaluated by using various techniques.The
Maxwell-Boltzmannian approach to nuclear reaction rate theory is
extended to cover Tsallis statistics (Tsallis, 1988) and more
general cases of distribution functions. The main purpose of this
paper is to investigate in some more detail the extended reaction
probability integral in the equilibrium thermodynamic argument and
in the cut-off case.  The extended reaction probability integrals
will be evaluated in closed form for all convenient values of the
parameter by means of residue calculus. A comparison of the standard
reaction probability integrals with the
extended reaction probability integrals is also done.\\

}\\
\begin{center}{\small {\section{\bf Introduction}}}\end{center}

The evolution of universe is due to the thermonuclear reactions which
 are taking place in hot cosmic plasma. The main concept behind
  the description of cosmic nucleosynthesis is the rate of nuclear
  reactions synthesizing light nuclei into heavier ones.  If we closely
  examine the nuclear cross section theoretically and experimentally, we
  can find the  analytical representations of thermonuclear reactions.
  Many researchers were looking for these analytic representations of
  the reaction rate probability integrals for the last few decades.
   Many approaches have been made on the study of thermonuclear
   reactions(Haubold and John, 1978; Haubold and Mathai, 1985,
    1998; Anderson et al, 1994, Saxena et al, 2004).  These studies will be effective only
     when the reaction probability integrals are expressed in computable
     series representations(Mathai and Haubold, 1988; Haubold and
     John, 1982).The derivations of closed-form representations of nuclear
     reaction rates and the approximations on them are based on the
     theory of generalized special functions.\\

     In the production of neutrinos in the gravitationally
     stabilized solar fusion reactor, due to the memory effects and
     long-range forces a possible deviation of the velocity
     distribution of plasma particles from Maxwell-Boltzmann is
     noted(Coraddu et al, 1999; Lavagno and Quarati, 2002; Lavagno and Quarati
     , 2006).  This was initiated by Tsallis' non-additive generalization
     of Boltzmann-Gibbs statistical Mechanics.  Tsallis statistics
     covers Boltzmann-Gibbs statistics(
     Tsallis, 1988; Gell-Mann and Tsallis, 2004; Tsallis, 2004).  The
     extension of the nuclear reaction rate theory from
     Maxwell-Boltzmann theory to Tsallis theory was done by Saxena
     et al (2004), Mathai (2005), Mathai and Haubold (2007).  In this
     scenario Mathai introduced a more general distribution
     function which can be incorporated in the reaction rate theory
     by appealing to entropic and distributional pathways. If Mathai's pathway model
     (Mathai, 2005; Mathai and Haubold, 2007) is introduced in the
       reaction probability integrals one goes into a wider class of integrals
       where the standard reaction probability integrals becomes a limiting
       case.\\

The paper is organized in the following way.  Section 2 contains the
standard representations of the non-resonant thermonuclear reaction
rates.  In section 3 we
 give the basic definitions that we use in our discussion.  Section 4 gives an outline
  of the extensions of the reaction probability integrals using the pathway model
   and also establish the series representations of the extended integrals
    $I_{1\alpha}$ and $I_{2\alpha}^{(d)}$.  Section 5 gives a comparison of the
    extended integrals with the standard integrals.
\begin{center}{\small{\section{\bf Standard representations of non-resonant
thermonuclear reaction rates}}}\end{center}

For the evaluation of the reaction rate $r_{ij}$ of
 the interacting particles $i$ and $j$ we have to consider the
 energies distributed between the particles.  The reacting particles in the
 astrophysical plasma follows a Maxwell-Boltzmann distribution.
  From Mathai and Haubold (1988) we can see that the expression for
   the reaction rate $r_{ij}$ of the reacting particles in the non-degenerate
   environment is
\begin{eqnarray}
r_{ij} &=& n_i n_j \left( \frac{8}{\pi \mu}\right)^{\frac{1}{2}}
{\left(\frac{1}{kT}\right)}^{\frac{3}{2}} \int _0^{\infty} E
\sigma(E) {\rm e}^{-\frac{E}{k T}} {\rm d} E \\
 &=& n_i n_j \langle {\sigma
\nu } \rangle \nonumber
\end{eqnarray}
where $n_i$ and $n_j$ are the number densities of the reacting
particles $ i$ and $j$, the reduced mass of the particles is denoted
by $\mu= \frac{m_i m_j}{m_i+m_j},~T $  is the temperature, $k$ is
the Boltzmann constant, the reaction cross section is $\sigma(E)$
and the kinetic energy of the particles in the center of mass system
is $E= \frac{\mu{\nu}^2}{2} $ where $\nu$ is the relative velocity
of the interacting particles $i$ and $j$.

We write $\langle {\sigma \nu } \rangle$ to indicate that it is an
appropriate average of the product
 of the cross section and relative velocity of the interacting particles.
  For detailed physical reasons see  Haubold and Mathai (1984, 1986).
{\subsection{\bf Standard non-resonant thermonuclear function}} When
two nuclei of charges $z_i$ and $z_j$ are colliding at low energies
below the coulomb barrier, the reaction cross section for the
non-resonant
 nuclear reactions have the form (Haubold and Mathai, 1998; Bergstroem et al, 1999,
 Mathai and Haubold, 2002)
\begin{eqnarray}
\sigma(E)&=& \frac{S(E)}{E}  {\rm e}^{-2 \pi \eta(E)}\\
\noalign{with} \eta(E)&=&\left(\frac{\mu}{2}\right)^{\frac{1}{2}}
\frac{z_i z_j {\rm e}^2}{\hbar E^{\frac{1}{2}}}
\end{eqnarray}
where $\eta(E) $ is the Sommerfeld parameter, $\hbar$ is the
Planck's quantum of action, ${\rm e}$ is the quantum of electric
charge, the cross section factor $S(E)$ is often found to be a
constant or a slowly varying function of energy over a limited range
of energy (Mathai and Haubold, 1988). The cross section factor
$S(E)$ can be parameterized by expanding in terms of the power
series about
 the zero energy because of its slow energy dependence.  $S(E)$ can be expressed as
\begin{equation}
 S(E)= S(0) + \frac{{\rm d}S(0)}{{\rm d}E}E +\frac{1}{2}
 \frac{{\rm d}^2S(0)}{{\rm d}E^2}E^2,
 \end{equation}
 where S(0) is the value of $S(E)$ at zero energy and $S'(0)$ and $S''(0)$ are
  the first and second order derivatives of $S(E)$ with respect to energy  evaluated
  at $ E=0,$  respectively.
 Then
 \begin{eqnarray}
  \langle {\sigma
\nu } \rangle  &=& \left( \frac{8}{\pi
\mu}\right)^{\frac{1}{2}}\sum_{\nu=0}^2{\left(\frac{1}{kT}\right)}^
{-\nu+\frac{1}{2}}\frac{S^{(\nu)}(0)}{\nu !}\times
\int_0^{\infty}E^\nu {\rm e}^{-\frac{E}{kT}-2 \pi \eta(E)}{\rm d}E \nonumber\\
   &=&\left( \frac{8}{\pi\mu}\right)^{\frac{1}{2}}
   \sum_{\nu=0}^2{\left(\frac{1}{kT}\right)}^{-\nu+\frac{1}{2}}
   \frac{S^{(\nu)}(0)}{\nu !}\times
\int_0^{\infty}x^\nu {\rm e}^{-x-bx^{-\frac{1}{2}}}{\rm d}x
 \end{eqnarray}
 where $x=\frac{E}{kT} $ and $b=\left(\frac{\mu}{2kT}\right)^{\frac{1}{2}}
\frac{z_i z_j {\rm e}^2}{\hbar }. $  The standard case of the thermonuclear
 function contains the nuclear cross section $\sigma(E)$, the energy dependent
  cross section factor $S(E)$ and the steady-state Maxwell-Boltzmann distribution function.

The collision probability integral for non-resonant thermonuclear
reactions in the Maxwell-Boltzmannian form is (Haubold and Mathai,
 1984)
\begin{equation}
I_1( \nu ,1 , b , \frac{1}{2} )=\int_0^{\infty}x^\nu {\rm
e}^{-x-bx^{-\frac{1}{2}}}{\rm d}x.
\end{equation}
We will consider the general integral
\begin{equation}
I_1(\gamma-1,a,b,\rho)= \int_0^{\infty}x^{\gamma-1} {\rm
e}^{-ax-bx^{-\rho}}{\rm d}x,~~ a>0, b>0, \rho>0.
\end{equation}
{\subsection{\bf Non-resonant thermonuclear function with high
energy cut-off}} It is assumed that the thermodynamic fusion plasma
is in exact thermodynamic equilibrium. But the cut-off of the high
energy tail of the Maxwell-Boltzmann
 distribution function in (2.6) results in a modification of the closed form
  representation of the appropriate quantity $\langle {\sigma \nu } \rangle$
  which is given by
\begin{equation}
I_2^{(d)}( \nu ,1 , b , \frac{1}{2} )=\int_0^d x^\nu {\rm
e}^{-x-bx^{-\frac{1}{2}}}{\rm d}x,~ b>0,~d<\infty.
\end{equation}
Again we consider the general form of the integral (2.8) as
\begin{equation}
I_2^{(d)}(\gamma-1,a,b,\rho)=\int_0^d x^{\gamma-1} {\rm
e}^{-ax-bx^{-\rho}}{\rm d}x,~ a>0,~ b>0,~ \rho>0,~d<\infty.
\end{equation}

For physical reasons for the cut-off modification of the
Maxwell-Boltzmann distribution function of the relative kinetic
energy of the reacting particles refer to the paper  Haubold and
Haubold and Mathai (1984).\\
{\subsection{\bf Modified non-resonant thermonuclear function with
depleted tail}}

A depletion of the high energy tail of the Maxwell-Boltzmann
distribution function of the relative kinetic energies of the nuclei
in the fusion plasma is explained in Haubold and Mathai(1986);
Haubold and John (1982); Kaniadakis et al (1997,1998).  The ad hoc
modification of the Maxwell-Boltzmann distribution for the
evaluation of the non-resonant thermonuclear reaction looks like a
depletion of the high energy tail of the Maxwell-Boltzmann
distribution. If their exists a possibility of such a modification a
remarkable change of the views of astrophysical nucleosynthesis and
controlled thermonuclear fusion may arise.

The integral form  $\langle {\sigma \nu } \rangle$  in comparison
with strict Maxwell-Boltzmannian case, we have the integral
\begin{equation}
I_3(\nu,1,\delta,b,\frac{1}{2}) =\int_0^{\infty}x^{\nu} {\rm
e}^{-x^{\delta} -bx^{-\frac{1}{2}}}{\rm d}x,~~b>0.
\end{equation}
We will consider the general integral of the type
\begin{equation}
I_3(\gamma-1, a , \delta , b ,\rho)= \int_0^{\infty}x^{\gamma-1}
{\rm e}^{-ax^\delta-bx^{-\rho}}{\rm d}x,
\end{equation}
where  $z>0,a>0, b>0, \rho>0.$\\
 \begin{center}{\small {\section{\bf Mathematical preliminaries}}}\end{center}
The basic quantities which we need in our discussion will be given
here. The gamma function denoted by $ \Gamma (z) $ for complex
number $z$ is defined as \begin {equation}
 \Gamma (z)= \int_0^{\infty} t^{z-1} {\rm e}^{-t} {\rm d} t, \Re (z)>0
 \end{equation}
where $\Re (\cdot)$ denotes the real part of $(\cdot)$. In general
$\Gamma (z)$ exists for all values of $z$, positive or negative,
except at the points $z= 0,-1,-2,\cdots $. These are the poles of
$\Gamma (z)$. But the integral representation holds for the real
part of $z$ to be positive. Another important result that we use in
our discussion is the psi function.  The psi function which is
denoted by $\psi(z)$  is the logarithmic derivative of a gamma
function and is defined as
\begin{equation}
\psi(z)=\frac{{\rm d}}{{\rm d}z} [\ln\Gamma(z)]=\frac{\frac{{\rm d}[\Gamma(z)]}
{{\rm d}z}}{\Gamma(z)} ~\mbox{or}~ \ln\Gamma(z)=\int_1^z \psi(x){\rm d}x.
\end{equation}
One property of the psi function that we will use is
\begin{equation}
\psi(1+n)=1+\frac{1}{2}+\frac{1}{3}+\cdots+\frac{1}{n}-\gamma
\end{equation}
where $\gamma$ is the Euler's constant and
$\gamma=0.5772156649\cdots,n=1,2,3,\cdots$.
\noindent
The G-function
which is originally due to C. S. Meijer in 1936 (See Mathai, 1993;
Mathai and Saxena,
 1973) is defined as a Mellin-Barnes type integral as follows:
\begin{equation}
 G_{p,q}^{m,n}\left(z \big|^{ a_1,\cdots,a_p}_{b_1,\cdots,b_q}\right)
=\frac{1}{2 \pi i}\int_L
\frac{\left\{\prod_{j=1} ^{m}\Gamma(b_j+s)\right\} \left\{ \prod_{j=1}
^{n}\Gamma(1-a_j-s)\right\}}{\left\{ \prod_{j=m+1} ^{q}\Gamma(1-b_j-s)\right\} \left\{
\prod_{j=n+1} ^{p}\Gamma(a_j+s)\right\}} z^{-s}{\rm d}s
\end{equation}
where $i=\sqrt{-1}$, $L$ is a suitable contour and $z\neq 0$,
$m,n,p,q$ are integers,  $0\leq m \leq q$ and $0\leq n\leq p$,
the empty product is interpreted as unity and the parameters $a_1,a_2,\cdots,a_p$ and
$b_1,b_2,\cdots,b_q$ are complex numbers such that no poles of
$\Gamma(b_j+s),~j=1,\cdots,m$ coincides with any pole of
$\Gamma(1-a_k-s),~k=1,\cdots,n;  $
\[-b_j-\nu \neq 1-a_k+\lambda,~~j=1,\cdots,m ;~~ k=1,\cdots,n;~~\nu,\lambda=0,1,\cdots. \]

This means that $a_k-b_j\neq\nu+\lambda+1$ or $a_k-b_j$ is not a
positive integer for$~~j=1,\cdots,m ;~ k=1,\cdots,n$.  We also
require that there is a strip in the complex $s$-plane which
separates the poles of $\Gamma(b_j+s),~ j=1,\cdots,m$ from those of
$\Gamma(1-a_k-s),~k=1,\cdots,n  $ (For the existance conditions and
properties of G-functions see Mathai (1993)).

Next we need the pathway model of Mathai (2005).  When fitting a
model to experimental data very often one needs a model with a
thicker or thinner tail than the ones available from a given
parametric family, or sometimes we may have a situation of the right
tail cut-off.  In order to take care of these situations and going
from one functional form to another, a pathway parameter is
introduced, see Mathai (2005) and Mathai and Haubold (2007).  By
this model we can proceed from a generalized type-1 beta model to a
generalized type-2 beta model to a generalized gamma model when the
variable is restricted to be positive. For the real scalar case the
pathway model is the following:
\begin{equation}
f(x)=c|x|^{\gamma-1}[1-a(1-\alpha)|x|^\delta]^{\frac{\eta}{1-\alpha}},~
a>0,\delta>0, 1-a(1-\alpha)|x|^\delta>0, \gamma>0, \eta>0
\end{equation}
 where $c$ is the
normalizing constant and $\alpha$ is the pathway parameter. When
$\alpha<1$ the model becomes a generalized type-1 beta model in the
real case.  This is a model with the right tail cut-off. When
$\alpha>1$ we have $1-\alpha=-(\alpha-1), \alpha>1$ so that
\begin{equation}
f(x)=c|x|^{\gamma-1}[1+a(\alpha-1)|x|^\delta]^{-\frac{\eta}{\alpha-1}},
\end{equation}
 which is a generalized type-2 beta model for real $x$. When
$\alpha\rightarrow1$ the above 2 forms will reduce to
\begin{equation}
f(x)=c|x|^{\gamma-1}{\rm e}^{-a\eta x^\delta}.
\end{equation}
Observe that the normalizing constant $c$ appearing in (3.5), (3.6)
and (3.7) are different.
\begin{center}
{\small {\section{\bf Extended thermonuclear function through
pathway model}}}
\end{center}

When Mathai's pathway model is introduced in the reaction
probability integrals we get a wider class of integrals.  Then the
standard reaction probability integrals
 become  particular cases of the new family of integrals.Through the pathway
 parameter $\alpha$
 we move to a wider class of integrals as $\alpha\rightarrow 1$ we get the
 reaction rate probability integrals.  If Maxwell-Boltzmann is the stable situation ,
  many unstable situations where Maxwell-Boltzmann is the limiting form are covered
   by the extended integrals.
   {\bf Extended integral in the standard non-resonant case}

   The extended integral in the
   standard non-resonant case is (Haubold and Kumar,
2007)
   \begin{equation}
   I_{1\alpha}=\int _{ 0 }^{\infty} x^{\gamma -1}[1+a(\alpha-1
)x]^{-\frac{1}{\alpha-1 }} {\rm e}^{-bx^{-\rho }}{\rm d}x.
   \end{equation}
\begin{thm}(Haubold and Kumar,
2007)
\begin{eqnarray}
I_{1\alpha}&=&\int _{ 0 }^{\infty} x^{\gamma -1}[1+a(\alpha-1
)x]^{-\frac{1}{\alpha-1 }}
{\rm e}^{-bx^{-\rho }}{\rm d}x  \nonumber \\
&=&\frac{1}{\rho [a(\alpha-1 )]^{\gamma }\Gamma \left
(\frac{1}{\alpha-1 } \right )} H_{1,2}^{2,1} \left(a(\alpha-1
)b^{\frac{1}{\rho }}\big| ^ {\left (1- \frac{1}{\alpha-1 }+\gamma ,
1 \right) }_ {(\gamma ,1),~(0,\frac{1}{\rho })} \right)
\end{eqnarray}
where $a>0,~b>0,~\rho>0,~\alpha>1,  ~\Re(s)>0,~ \Re(\gamma+s)>0$.
\end{thm}
When $\alpha\rightarrow1,~I_{1\alpha}$ becomes $I_{1}$. But $
I_{1\alpha} $ contains all neighborhood solutions for various values
of
$\alpha$  for $\alpha>1$.\\
If in the above result  $\frac{1}{\rho}$ is an integer then by
taking  $ \frac{1}{\rho}= m $ we obtain
\begin{cor}.
For $a>0,~b>0$ and $\alpha>1$, we have
\begin{eqnarray}
&& \int _{ 0 }^{\infty} x^{\gamma -1}[1+a(\alpha-1
)x]^{-\frac{1}{\alpha-1 }}
{\rm e}^{-bx^{-\frac{1}{m}}}{\rm d}x \nonumber \\
&&=\frac{\sqrt{m}(2 \pi)^{\frac{1-m}{2}}}{ [a(\alpha-1 )]^{\gamma
}\Gamma \left (\frac{1}{\alpha-1 } \right )}G_{1,m+1}^{m+1,1}
\left(\frac{a(\alpha-1)b^m}{m^m} \big|^{1-\frac{1}{\alpha-1 }+\gamma
}_{ 0,\frac{1}{m},\cdots,\frac{m-1}{m},\gamma} \right).
\end{eqnarray}
\end{cor}
By setting $\gamma-1=\nu,~a=1$ and $\rho =\frac{1}{2}$, we obtain
\begin{cor}. For $b>0,~\alpha>1$
 \begin{eqnarray}
&&\int _{ 0 }^{\infty} x^{\nu}[1+(\alpha-1 )x]^{-\frac{1}{\alpha-1
}}
{\rm e}^{-bx^{-\frac{1}{2} }}{\rm d}x  \nonumber \\
&&= \frac{(\pi)^{-\frac{1}{2}}}{(\alpha-1 )^{\nu+1 }\Gamma \left
(\frac{1}{\alpha-1 } \right )}G_{1,3}^{3,1} \left(
\frac{(\alpha-1)b^2}{4} \big|^{2-\frac{1}{\alpha-1}+\nu}_
 {0, \frac{1}{2}, \nu+1} \right).
 \end{eqnarray}
 \end{cor}
 {\bf Extended cut-off case}\\
 In the case of non-resonant thermonuclear reactions with high energy
cut-off the extended integral is (Haubold and
 Kumar, 2007)
 \begin{eqnarray*}
 I_{2\alpha}^{(d)}&=&\int _{ 0 }^{d} x^{\gamma -1}[1-a(1-\alpha
)x]^{\frac{1} {1-\alpha }}{\rm e}^{-bx^{-\rho }}{\rm d}x
\end{eqnarray*}
where $ d=\frac{1}{a(1-\alpha )},~\alpha <1,~a>0,~\delta
=1,~\eta=1,~1-a(1-\alpha )x>0,~\rho >0,~b>0$.
\begin{thm}
\begin{eqnarray}
&&\int _{ 0 }^{d} x^{\gamma -1}[1-a(1-\alpha )x]^{\frac{1}{1-\alpha
}}
{\rm e}^{-bx^{-\rho }}{\rm d}x  \\
&&=\frac{\Gamma \left ( \frac{1}{1-\alpha }+1 \right )} {\rho
[a(1-\alpha )]^{\gamma }} H_{1,2}^{2,0} \left(a(1-\alpha )
b^{\frac{1}{\rho }}\big|^ {\left ( 1+\gamma  + \frac{1}{1-\alpha } ,
1 \right) }_{ (\gamma ,1),~(0,\frac{1}{\rho })}
\right)=I_{2\alpha}^{(d)}
\end{eqnarray}
where $a>0,~ b>0,~ \rho>0,~ \alpha <1,~\Re(\gamma+s)>0$ and
$d<\infty$
\end{thm}
When $\alpha\rightarrow1,~I_{2\alpha}^{(d)}$ becomes $I_{2}^{(d)}$.
But $ I_{2\alpha}^{(d)} $ contains all neighborhood solutions for
various values of $\alpha$  for $\alpha<1$. In the above result if
$\frac{1}{\rho}$ is an integer then by taking  $ \frac{1}{\rho}= m $
we obtain
\begin{cor}.
For $a>0,~ b>0,~ \alpha <1,~d<\infty$ \text{and} $\Re(\gamma+s)>0$
\begin{eqnarray}
&&\int _{ 0 }^{d} x^{\gamma -1}[1-a(1-\alpha )x]^{\frac{1}{1-\alpha
}}
{\rm e}^{-bx^{-\frac{1}{m}}}{\rm d}x  \nonumber\\
&&=\frac{\sqrt{m} {(2 \pi )^{\frac{1-m}{2}}} \Gamma \left (
\frac{1}{1-\alpha }+1 \right )}{ [a(1-\alpha )]^{\gamma }}
G_{1,m+1}^{m+1,0} \left(\frac{a(1-\alpha)b^m}{m^{m}} \big|^{1+\gamma
+\frac{1}{1-\alpha }}_ {0,\frac{1}{m},\cdots,\frac{m-1}{m},\gamma}
\right)
\end{eqnarray}
\end{cor}
By setting $\gamma-1=\nu,~a=1$ and $\rho =\frac{1}{2}$, we obtain
\begin{cor}.
For $b>0,~ \alpha <1,~d<\infty$ \text{and} $\Re(\nu+1+s)>0$
\begin{eqnarray}
&&\int _{ 0 }^{d} x^{\nu}[1-(1-\alpha )x]^{\frac{1}{1-\alpha }}
{\rm e}^{-bx^{-\frac{1}{2}}}{\rm d}x  \nonumber\\
 &&=\frac{\Gamma \left ( \frac{1}{1-\alpha }+1 \right )}
{\sqrt{\pi} (1-\alpha )^{\nu+1 }} G_{1,3}^{3,0} \left(
\frac{(1-\alpha )b^2}{4} \big|^{\nu +\frac{1}{1-\alpha }+2}_
 {0,\frac{1}{2},\nu+1} \right)
\end{eqnarray}
\end{cor}
{\bf Extended depleted case}\\
\noindent
 Proceeding
similarly as in the case of $I_{1\alpha}$ we get for the depleted
case
\begin{thm}
\begin{eqnarray}
I_{3\alpha}&=&\int _{ 0 }^{\infty} x^{\gamma -1}[1+a(\alpha-1
)x^\delta]^{-\frac{1}{\alpha-1 }}
{\rm e}^{-bx^{-\rho }}{\rm d}x \nonumber\\
&&=\frac{1}{\rho [a(\alpha-1 )]^{\frac{\gamma}{\delta} }\Gamma \left
(\frac{1}{\alpha-1 } \right)} H_{1,2}^{2,1} \left([a(\alpha-1
)]^{\frac{1}{\delta}}b^{\frac{1}{\rho }}\big| ^ {\left(1-
\frac{1}{\alpha-1 }+\frac{\gamma}{\delta} , \frac{1}{\delta}
\right)}_ {(\frac{\gamma}{\delta}
,\frac{1}{\delta}),~(0,\frac{1}{\rho })} \right)
\end{eqnarray}
where $a>0,~b>0,~\rho>0,~\delta>0,~\alpha>1,  ~\Re(s)>0,~
\Re(\gamma+s)>0$.
\end{thm}

 For the non-resonant thermonuclear reactions with
depleted tail $\gamma-1=\nu,~a=1,~\rho =\frac{1}{2}$ then we get,
\begin{cor}
\begin{eqnarray}
&&\int _{ 0 }^{\infty} x^{\nu}[1+(\alpha-1
)x^\delta]^{-\frac{1}{\alpha-1 }}
{\rm e}^{-bx^{-\frac{1}{2} }}{\rm d}x \nonumber\\
&=&\frac{2}{\delta(\alpha-1 )^{\frac{\nu+1}{\delta} }\Gamma \left
(\frac{1}{\alpha-1 } \right )} H_{1,2}^{2,1} \left((\alpha-1
)^{\frac{1}{\delta}}b^2\big| ^ {\left (1-\frac{1}{\alpha-1
}+\frac{\nu+1}{\delta} , \frac{1}{\delta} \right)}_
{(\frac{\nu+1}{\delta} ,\frac{1}{\delta}),~(0,2)} \right).
\end{eqnarray}
\end{cor}
By setting $\frac{s}{\delta}=s'$ and $2\delta=m,~m=1,2,\cdots$ we
get
\begin{cor}
\begin{eqnarray}
I_{3\alpha}
&=&\frac{2(2\pi)^{\frac{1-m}{2}}m^{-\frac{1}{2}}}{(\alpha-1
)^{\frac{2(\nu+1)}{m} }\Gamma \left (\frac{1}{\alpha-1 } \right )}
G_{1,m+1}^{m+1,1} \left(\frac{(\alpha-1 )b^m}{m^m}\big| ^ {1-
\frac{1}{\alpha-1 }+\frac{2(1+\nu)}{m}}_ {0,\frac{1}{m },\frac{2}{m
},\cdots,\frac{m-1}{m },\frac{2(1+\nu)}{m}} \right)
\end{eqnarray}
\end{cor}
{\subsection{\bf Series representations}}

In the following we derive series representations of the right-hand
side of (4.4) which will be helpful for the evaluation of the
extended reaction probability integrals in the Maxwell-Boltzmannian
form. Taking $\nu$ as a general parameter one can consider several
situations.  Then following through the process in the papers of
Haubold and Mathai, see for example Mathai and Haubold (1988), we
have the following series representations for the extended integral
in (4.4).\\
 {\subsubsection{\bf Case (I): $\nu \neq
\pm\frac{\lambda}{2},~ \lambda=0,1,2,\cdots$ } } Here we apply
residue calculus on the G-function for obtaining the series
 representation of the integrals. Consider the G-function in (4.4).
 \begin{eqnarray}
&&G_{1,3}^{3,1} \left( \frac{(\alpha-1)b^2}{4}
\big|^{2-\frac{1}{\alpha-1}+\nu}_
 {0, \frac{1}{2}, \nu+1} \right) = \frac{1}{2 \pi i}
 \int
_{c-i{\infty}}^{c+i{\infty}}\Gamma(s)\Gamma \left(
\frac{1}{2}+s\right)\nonumber \\
&&\times  \Gamma (1+\nu+s )\Gamma \left ( \frac{1}{\alpha-1 }-\nu-1-s \right)
  \left( \frac{(\alpha-1)b^2}{4} \right)^{-s} {\rm d}s
 \end{eqnarray}
\noindent The right hand side is the sum of the residues of the
integrand. The poles of the gammas in the
integral representation in (4.12) are as follows.\\
\medskip
Poles of $\Gamma(s) : ~ s=0,-1,-2,\cdots;~\Gamma \left(
\frac{1}{2}+s\right) :~ s =
-\frac{1}{2},-\frac{3}{2},-\frac{5}{2},\cdots ;~\Gamma (1+\nu+s ):~s=-\nu-1,-\nu-2,-\nu-3,\cdots.$ \\
These are all simple poles under the conditions in case(1). Then the
G-function has a simple series expansion.

 \noindent
 We know that\\
\begin{eqnarray}
\lim_{s\rightarrow -r}
(s+r)\Gamma(s)&=&\frac{(-1)^r}{r!},\\
\Gamma(a-r)&=&\frac{(-1)^r\Gamma(a)}{(1-a)_r},\\
\Gamma(a+m)&=&\Gamma(a)(a)_m
\end{eqnarray}
when $\Gamma(a)$ is defined,
  $r=0,1,2,\cdots ;~\Gamma
\left(\frac{1}{2}\right)={\pi}^{\frac{1}{2}},$
\\
$(a)_r=\left\{ \begin{array}{ll}
a(a+1)\cdots(a+r-1)& \text{if}~r\geq1,~a\neq0\\
1 & \text{if}~ r=0,\end{array}\right.$\\
The sum of the residues corresponding to the poles
 $s=-r,r=0,1,2,\cdots$ is given by
\begin{eqnarray}
R_1&=&\sum_{r=0}^{\infty}\frac{(-1)^r}{r!}\Gamma\left(\frac{1}{2}-r\right)
\Gamma\left(\frac{1}{\alpha-1}-1-\nu+r\right)
\Gamma(1+\nu-r)\left[\frac{(\alpha-1)b^2}{4}\right]^r\nonumber\\
&=&{\pi}^{\frac{1}{2}}\Gamma\left(\frac{1}{\alpha-1}-1-\nu\right)
\Gamma(1+\nu){_1F_2}
\bigg(\frac{1}{\alpha-1}-1-\nu;\frac{1}{2},-\nu;-\frac{(\alpha-1)b^2}{4}\bigg)\nonumber \\
\end{eqnarray}
where ${_pF_q}(a_p;b_q;z)$ denotes the generalized hypergeometric
 function defined as above.

The sum of the residues corresponding to the poles $ s =
-\frac{1}{2},-\frac{3}{2},-\frac{5}{2},\cdots $ is
\begin{eqnarray}
R_2&=&-2{\pi}^{\frac{1}{2}}\Gamma\left(\frac{1}{\alpha-1}-\frac{1}{2}-\nu\right)
\Gamma\left(\frac{1}{2}+\nu\right)\left[\frac{(\alpha-1)b^2}{4}\right]^
{\frac{1}{2}}\nonumber\\
&&\times
{_1F_2}\left(\frac{1}{\alpha-1}-\frac{1}{2}-\nu;~\frac{3}{2},\frac{1}{2}-\nu;
~-\frac{(\alpha-1)b^2}{4}\right).
\end{eqnarray}
Finally the sum of the residues corresponding to
$s=-\nu-1,-\nu-2,-\nu-3,\cdots$ is
\begin{eqnarray}
R_3&=&\Gamma(-\nu-1)\Gamma\left(-\nu-\frac{1}{2}\right)\Gamma\left(\frac{1}{\alpha-1}\right)
\left[\frac{(\alpha-1)b^2}{4}\right]^{1+\nu} \nonumber \\
&&\times {_1F_2}\bigg(\frac{1}{\alpha-1};2+\nu,
\frac{3}{2}+\nu;-\frac{(\alpha-1)b^2}{4}\bigg).
\end{eqnarray}
Adding $R_1, R_2, R_3$ we obtain the final result:
\begin{thm}
If $\nu \neq \pm\frac{\lambda}{2},~ \lambda=0,1,2,\cdots$ is an
integer, then for $b>0,~\alpha>1$, we have
 \begin{eqnarray}
 \int _{ 0 }^{\infty} x^{\nu}[1+(\alpha-1
)x]^{-\frac{1}{\alpha-1 }} {\rm e}^{-bx^{-\frac{1}{2} }}{\rm d}x
&=&\bigg\{\frac{\Gamma\left(\frac{1}{\alpha-1}-1-\nu\right)
\Gamma(1+\nu)}{[(\alpha-1 )]^{\nu+1 }\Gamma \left (\frac{1}{\alpha-1
} \right )}\nonumber \\
&&\times {_1F_2}
\bigg(\frac{1}{\alpha-1}-1-\nu;\frac{1}{2},-\nu;-\frac{(\alpha-1)b^2}
{4}\bigg)\bigg\}\nonumber\\
&&-\bigg\{\frac{2\Gamma\left(\frac{1}
{\alpha-1}-\frac{1}{2}-\nu\right)
\Gamma\left(\frac{1}{2}+\nu\right)}{[(\alpha-1 )]^{\nu+1 }\Gamma \left
(\frac{1}{\alpha-1 } \right
)}\left[\frac{(\alpha-1)b^2}{4}\right]^{\frac{1}{2}}
\nonumber\\
&&\times
{_1F_2}\left(\frac{1}{\alpha-1}-\frac{1}{2}-\nu;~
\frac{3}{2},\frac{1}{2}-\nu;~-\frac{(\alpha-1)b^2}{4}\right)\bigg\}\nonumber\\
&&+\bigg\{\frac{(\pi)^{-\frac{1}{2}}\Gamma(-\nu-1)
\Gamma\left(-\nu-\frac{1}{2}\right)}{[(\alpha-1 )]^{\nu+1 }}
\bigg[\frac{(\alpha-1)b^2}{4}\bigg]^{1+\nu} \nonumber \\
&&\times
{_1F_2}\bigg(\frac{1}{\alpha-1};2+\nu,\frac{3}{2}+\nu;-\frac{(\alpha-1)b^2}
{4}\bigg)\bigg\}
\end{eqnarray}
\end{thm}
It is to be noted that the series on the right-hand side of the
equation (4.19) are term-wise integrable over any finite range.
{\subsubsection{ \bf Case (II): $\nu$ is a positive integer}} In
this case some poles of $\Gamma(s)$ and $\Gamma (1+\nu+s )$ will
coincide with each other. Therefore these will be of order 2.  We
note that the poles $s=0,-1,-2,\cdots,-\nu $ are each of order 1: $s
= -\frac{1}{2},-\frac{3}{2},-\frac{5}{2},\cdots $  are each of order
1: $s=-\nu-1,-\nu-2,-\nu-3,\cdots$ are each of order 2. Taking the
sum of residues at the poles $s=0,-1,-2,\cdots,-\nu$; at $ s =
-\frac{1}{2},-\frac{3}{2},-\frac{5}{2},\cdots$; at
$s=-\nu-1,-\nu-2,-\nu-3,\cdots$ we have
\begin{equation}
R_1={\pi}^{\frac{1}{2}}\Gamma\left(\frac{1}{\alpha-1}-1-\nu\right)
\Gamma(1+\nu)\sum_{r=0}^{\nu}\frac{(-1)^r}{r!}
\frac{\left(\frac{1}{\alpha-1}-1-\nu\right)_r}
{\left(\frac{1}{2}\right)_r(-\nu)_r}\left[\frac{(\alpha-1)b^2}{4}\right]^r
\end{equation}
\begin{eqnarray}
R_2&=&-2{\pi}^{\frac{1}{2}}\Gamma\left(\frac{1}{\alpha-1}-\frac{1}{2}-\nu\right)
\Gamma\left(\frac{1}{2}+\nu\right)\left[\frac{(\alpha-1)b^2}{4}
\right]^{\frac{1}{2}}\nonumber\\
&&\times
{_1F_2}\left(\frac{1}{\alpha-1}-\frac{1}{2}-\nu;~\frac{3}{2},\frac{1}{2}-\nu;
~-\frac{(\alpha-1)b^2}{4}\right)
\end{eqnarray}
\begin{eqnarray}
R_3=\left(\frac{(\alpha-1)b^2}{4}\right)^{1+\nu}\sum_{r=0}^{\infty}
\left(\frac{(\alpha-1)b^2}{4}\right)^r \bigg[-\ln\left(
\frac{(\alpha-1)b^2}{4}\right)+A_r\bigg] B_r,
\end{eqnarray}
where
\begin{equation}
A_r=\psi\left(-\frac{1}{2}-\nu-r\right)+
\psi\left(\frac{1}{\alpha-1}+r \right)+\psi(1+r)
+\psi(2+\nu+r)
\end{equation}
and
\begin{equation}
B_r=\frac{(-1)^{1+\nu +r}\Gamma\left(-\frac{1}{2}-\nu \right)
\Gamma\left(\frac{1}{\alpha-1}\right)
\left(\frac{1}{\alpha-1}\right)_r}
{\left(\frac{3}{2}+\nu \right)_r r!(1+\nu +r)!}
\end{equation}

From the above results we have the following theorem.
\begin{thm}
If $\nu> 0$ is an integer, then for $b>0,~\alpha>1$, we have
 \begin{eqnarray}
&&\int _{ 0 }^{\infty} x^{\nu}[1+(\alpha-1 )x]^{-\frac{1}{\alpha-1
}} {\rm e}^{-bx^{-\frac{1}{2} }}{\rm d}x\nonumber \\
&=&\frac{(\pi)^{-\frac{1}{2}}}{[(\alpha-1 )]^{\nu+1 }\Gamma \left
(\frac{1}{\alpha-1 } \right
)}\bigg\{{\pi}^{\frac{1}{2}}\Gamma\left(\frac{1}{\alpha-1}-1-\nu\right)
 \Gamma(1+\nu) \nonumber \\
 && \times \sum_{r=0}^{\nu}\frac{(-1)^r}{r!}
\frac{\left(\frac{1}{\alpha-1}-1-\nu\right)_r}
{\left(\frac{1}{2}\right)_r(-\nu)_r}\left[\frac{(\alpha-1)b^2}{4}\right]^r \nonumber \\
&&-2{\pi}^{\frac{1}{2}}\Gamma\left(\frac{1}{\alpha-1}-\frac{1}{2}-\nu\right)
\Gamma\left(\frac{1}{2}+\nu\right)\left[\frac{(\alpha-1)b^2}{4}\right]^{\frac{1}{2}}
\nonumber \\
&&\times
{_1F_2}\left(\frac{1}{\alpha-1}-\frac{1}{2}-\nu;~\frac{3}{2},\frac{1}{2}-\nu;
~-\frac{(\alpha-1)b^2}{4}\right) \nonumber \\
&&+\left(\frac{(\alpha-1)b^2}{4}\right)^{1+\nu}\sum_{r=0}^{\infty}
\left(\frac{(\alpha-1)b^2}{4}\right)^r \bigg[-\ln\left(
\frac{(\alpha-1)b^2}{4}\right)+A_r\bigg] B_r\bigg\}
\end{eqnarray}
where $A_r$ and $B_r$ are as given in (4.23) and (4.24) .
\end{thm}
{\subsubsection{\bf{Case (III): $\nu$ a negative integer}} Let
$\nu=-\mu,\mu>0.$  Then the poles of the G-function in (4.12) are $s
= -\frac{1}{2},-\frac{3}{2},-\frac{5}{2},\cdots $  are each of order
1; $s=-\nu-1,-\nu-2,-\nu-3,\cdots,+1$ are each of order 1; poles
$s=0,-1,-2,\cdots $ are each of order 2.Again, following through the
same stages as above we have
\begin{thm}
For $\nu$ a negative integer,
\begin{eqnarray}
&&\int _{ 0 }^{\infty} x^{\nu}[1+(\alpha-1 )x]^{-\frac{1}{\alpha-1
}} {\rm e}^{-bx^{-\frac{1}{2} }}{\rm d}x\nonumber \\
&=&\frac{(\pi)^{-\frac{1}{2}}}{(\alpha-1 )^{\nu+1 }\Gamma \left
(\frac{1}{\alpha-1 } \right
)}\bigg\{-2{\pi}^{\frac{1}{2}}\Gamma\left(\frac{1}{\alpha-1}-\frac{1}{2}-\nu\right)
\Gamma\left(\frac{1}{2}+\nu\right)\left[\frac{(\alpha-1)b^2}{4}
\right]^{\frac{1}{2}}\nonumber\\
&&\times
{_1F_2}\left(\frac{1}{\alpha-1}-\frac{1}{2}-\nu;~\frac{3}{2},\frac{1}{2}-\nu;
~-\frac{(\alpha-1)b^2}{4}\right)\nonumber \\
&+&\Gamma(-1-\nu)\Gamma\left(-\frac{1}{2}-\nu\right)\Gamma\left(\frac{1}{\alpha-1}\right)
\left[\frac{(\alpha-1)b^2}{4}\right]^{1+\nu} \nonumber \\
&&\times \sum_{r=0}^{-\nu-2}\frac{(-1)^r}{r!}
\frac{\left(\frac{1}{\alpha-1}\right)_r}
{\left(\frac{3}{2}+\nu\right)_r(2+\nu)_r}\left[\frac{(\alpha-1)b^2}{4}\right]^r
\nonumber \\
&+&\sum_{r=0}^{\infty} \left(\frac{(\alpha-1)b^2}{4}\right)^r
\bigg[-\ln\left( \frac{(\alpha-1)b^2}{4}\right)+{A'}_r\bigg]
{B'}_r\bigg\},
\end{eqnarray}
where
\begin{equation}
{A'}_r=\psi\left(\frac{1}{2}-r\right)+
\psi\left(\frac{1}{\alpha-1}-1-\nu+r \right)+\psi(1+r) +\psi(r-\nu)
\end{equation}
and
\begin{equation}
{B'}_r=\frac{(-1)^{1+\nu }\Gamma\left(\frac{1}{2}-r \right)
\Gamma\left(\frac{1}{\alpha-1}-1-\nu-r\right) } { r!(r-\nu-1)!}
\end{equation}
\end{thm}

 {\subsubsection{\bf{Case (IV): $\nu$ a
positive half integer}} Let $\nu=m+\frac{1}{2},m=0,1,2,\cdots.$ Then
\begin{eqnarray*}
\Gamma(s)\Gamma \left( \frac{1}{2}+s\right)
\Gamma (1+\nu+s )\Gamma \left ( \frac{1}{\alpha-1 }-\nu-1-s \right)\\
=\Gamma(s)\Gamma \left( \frac{1}{2}+s\right)
  \Gamma (m+\frac{3}{2}+s )\Gamma \left ( \frac{1}{\alpha-1 }-m-\frac{3}{2}-s \right)
  \end{eqnarray*}
Here the poles of $\Gamma(\frac{1}{2}+s)$ and
$\Gamma(s+m+\frac{3}{2})$ coincides with each other.  These will be
of order 2.  We note that the poles $s=0,-1,-2,\cdots$ are each of
order 1; $s
=-\frac{1}{2},-\frac{1}{2}-1,-\frac{1}{2}-2,\cdots,-\frac{1}{2}-m $
are each of order 1;$s=-m-\frac{3}{2},-m-\frac{5}{2},\cdots $ are
each of order 2 and we have the following theorem
\begin{thm}
For $\nu$ a positive half integer, namely $\nu=m+\frac{1}{2}, ~m=0,1,2,\cdots$
\begin{eqnarray}
&&\int _{ 0 }^{\infty} x^{\nu}[1+(\alpha-1 )x]^{-\frac{1}{\alpha-1
}} {\rm e}^{-bx^{-\frac{1}{2} }}{\rm d}x \nonumber \\
&=& \frac{(\pi)^{-\frac{1}{2}}}{(\alpha-1 )^{\frac{3}{2}+m }\Gamma
\left (\frac{1}{\alpha-1 } \right )}\bigg\{{\pi}^{\frac{1}{2}}
\Gamma\left(\frac{1}{\alpha-1}-\frac{3}{2}-m\right)
\Gamma\left(\frac{3}{2}+m\right) \nonumber \\
&&\times{_1F_2}\left( \frac{1}{\alpha-1}-\frac{3}{2}-m;\frac{1}{2},
-\frac{1}{2}-m;-\frac{(\alpha-1)b^2}{4}\right) \nonumber \\
&-&2{\pi}^{\frac{1}{2}}\Gamma\left(\frac{1}{\alpha-1}-1-m\right)
\Gamma(1+m)\left[\frac{(\alpha-1)b^2}{4}\right]^{\frac{1}{2}} \nonumber \\
&\times& \sum_{r=0}^{m}\frac{(-1)^r}{r!}\frac{\left(\frac{1}
{\alpha-1}-1-m\right)_r}{\left(\frac{3}{2}\right)_r
(-m)_r}\left[\frac{(\alpha-1)b^2}{4}\right]^{r} \nonumber \\
&+&\left(\frac{(\alpha-1)b^2}{4}\right)^{m+\frac{3}{2}}\sum_{r=0}^{\infty}
\left(\frac{(\alpha-1)b^2}{4}\right)^r \bigg[-\ln\left(
\frac{(\alpha-1)b^2}{4}\right)+C_r\bigg]D_r\bigg\},
\end{eqnarray}
where
\begin{equation}
C_r=\psi\left(-m-\frac{3}{2}-r\right)+
\psi\left(\frac{1}{\alpha-1}+r \right)+\psi(1+r) +\psi(2+m+r)
\end{equation}
and
\begin{equation}
D_r=\frac{(-1)^{1+m+r}\Gamma\left(-m-\frac{3}{2} \right)
\Gamma\left(\frac{1}{\alpha-1}\right)
\left(\frac{1}{\alpha-1}\right)_r} {\left(\frac{5}{2}+m \right)_r
r!(1+m +r)!}
\end{equation}
\end{thm}
 {\subsubsection{\bf{Case (V): $\nu$ a
negative half integer}} Let $\nu=-m-\frac{1}{2},m=0,1,2,\cdots.$
Then in this case
\begin{eqnarray*}
\Gamma(s)\Gamma \left( \frac{1}{2}+s\right)
\Gamma (1+\nu+s )\Gamma \left ( \frac{1}{\alpha-1 }-\nu-1-s \right)\\
=\Gamma(s)\Gamma \left( \frac{1}{2}+s\right)
  \Gamma (s+\frac{1}{2}-m )\Gamma \left ( \frac{1}{\alpha-1 }-\frac{1}{2}+m-s \right)
  \end{eqnarray*}
  Then the poles of Meijer's G-function in (4.12) are
$s=-r,~ r=0,1,2,\cdots $ of order 1 each;  $ s=m-\frac{1}{2}-r,~
r=0,1,2,\cdots,m-1 $ of order 1 each;  $ s=-\frac{1}{2}-r,~
r=0,1,2,\cdots $of order 2 each. Then we have the following theorem
\begin{thm}
For $\nu$ a negative half integer, namely $\nu=-m-\frac{1}{2},
~m=0,1,2,\cdots$
\begin{eqnarray}
&&\int _{ 0 }^{\infty} x^{\nu}[1+(\alpha-1 )x]^{-\frac{1}{\alpha-1
}} {\rm e}^{-bx^{-\frac{1}{2} }}{\rm d}x\nonumber
\\
 &=& \frac{(\pi)^{-\frac{1}{2}}}{(\alpha-1 )^{\frac{3}{2}+m }\Gamma
\left (\frac{1}{\alpha-1 } \right
)}\bigg\{\sqrt{\pi}\Gamma\left(\frac{1}{2}-m\right)\Gamma\left(
\frac{1}{\alpha-1 }-\frac{1}{2}+m \right)\nonumber
\\
&&\times{_1F_2}\left(\frac{1}{\alpha-1 }-\frac{1}{2}+m
;\frac{1}{2},\frac{1}{2}+m;-\frac{(\alpha-1)b^2}{4}\right)\nonumber
\\
&&+\Gamma\left(m-\frac{1}{2}\right)\Gamma(m)\Gamma\left(\frac{1}{\alpha-1
}\right)\left(\frac{(\alpha-1)b^2}{4}\right)^{-m+\frac{1}{2}}\nonumber
\\
&&\times \sum_{r=0}^{m}\frac{(-1)^r}{r!}\frac{\left(\frac{1}
{\alpha-1}\right)_r}{\left(\frac{3}{2}-m\right)_r
(1-m)_r}\left[\frac{(\alpha-1)b^2}{4}\right]^{r}\nonumber \\
&&+\left(\frac{(\alpha-1)b^2}{4}\right)^{\frac{1}{2}}\sum_{r=0}^{\infty}
\left(\frac{(\alpha-1)b^2}{4}\right)^r \bigg[-\ln\left(
\frac{(\alpha-1)b^2}{4}\right)+{C'}_r\bigg]{D'}_r\bigg\},
\end{eqnarray}
where
\begin{equation}
{C'}_r=\psi\left(-\frac{1}{2}-r\right)+
\psi\left(\frac{1}{\alpha-1}+m+r \right)+\psi(1+r) +\psi(1+m+r)
\end{equation}
and
\begin{equation}
{D'}_r=\frac{\sqrt{\pi}(-1)^{m+r}
\Gamma\left(\frac{1}{\alpha-1}+m+r\right)
\left(\frac{1}{\alpha-1}\right)_r} {\left(\frac{3}{2}\right)_r r!(m
+r)!}.
\end{equation}
\end{thm}
 {\subsection{\bf Series representation for the extended cut-off case}}

 {\subsubsection{\bf Case(I): $\nu
\neq \pm\frac{\lambda}{2},~ \lambda=0,1,2,\cdots$ }} Here we apply
the same techniques that we applied previously.  Consider the
evaluation of the G-function in (4.8).
\begin{eqnarray}
&&\int _{ 0 }^{d} x^{\nu}[1-(1-\alpha )x]^{\frac{1}{1-\alpha }}
{\rm e}^{-bx^{-\frac{1}{2}}}{\rm d}x  \nonumber\\
&=&\frac{\Gamma \left ( \frac{1}{1-\alpha }+1 \right )}{\sqrt{\pi}
(1-\alpha )^{\nu+1 }} \frac{1}{2 \pi i} \int_{c-i \infty }^{c+ i
\infty } \frac{ \Gamma (s) \Gamma \left( s+\frac{1}{2}\right)\Gamma
(1+\nu+s)
 }{\Gamma \left (2+\nu+\frac{1}{1-\alpha } +s \right )}
 \left[ \frac{(1-\alpha )b^2}{4}\right]^{-s} {\rm d} s
 \end{eqnarray}

 The poles of the integral
  are as follows:
  $\Gamma(s) : ~ s=0,-1,-2,\cdots; ~\Gamma \left( \frac{1}{2}+s\right) :~ s =
-\frac{1}{2},-\frac{3}{2},-\frac{5}{2},\cdots ;~\Gamma (1+\nu+s ):
 ~s=-\nu-1,-\nu-2,-\nu-3,\cdots.$ \\
These are all simple poles under case (1). Then evaluating the sum
of residues we have
\begin{thm}
If $\nu \neq \pm \frac{\lambda}{2},~ \lambda=0,1,2,\cdots$ is an
integer,  then for $b>0,~\alpha<1$, we have
 \begin{eqnarray}
 &&\int _{ 0 }^{d} x^{\nu}[1-(1-\alpha )x]^{\frac{1}{1-\alpha }}
{\rm e}^{-bx^{-\frac{1}{2}}}{\rm d}x  \nonumber\\
 &=&\frac{\Gamma \left ( \frac{1}{1-\alpha }+1 \right )}
{\sqrt{\pi} (1-\alpha )^{\nu+1 }}\bigg\{
\frac{(\pi)^{\frac{1}{2}}\Gamma(1+\nu)}{\Gamma \left
(2+\nu+\frac{1}{1-\alpha }
 \right ) }{_1F_2}\left(-1-\nu-\frac{1}{1-\alpha};~\frac{1}{2},~-\nu;
 ~\frac{(1-\alpha )b^2}{4}\right)  \nonumber \\
&-&\frac{2(\pi)^{\frac{1}{2}}\Gamma\left(\frac{1}{2}+\nu\right)}{\Gamma\left
(\frac{3}{2}+\nu+\frac{1}{1-\alpha }
 \right )}\left( \frac{(1-\alpha )b^2}{4}\right)^{\frac{1}{2}}
{_1F_2}\left(-\frac{1}{2}-\nu-\frac{1}{\alpha-1};~\frac{3}{2},\frac{1}{2}-\nu;~
\frac{(1-\alpha )b^2}{4}\right) \nonumber\\
&+&\frac{\Gamma(-\nu-1)\Gamma\left(-\nu-\frac{1}{2}\right)}{\Gamma\left
(1+\frac{1}{1-\alpha }
 \right )}\left( \frac{(1-\alpha )b^2}{4}\right)^{1+\nu} {_1F_2}\left(1+\frac{1}{1-\alpha};~2+\nu,\frac{3}{2}+\nu;~
-\frac{(1-\alpha )b^2}{4}\right)\bigg\}\nonumber \\
\end{eqnarray}
\end{thm}
Here the series on the right-hand side of the equation (4.36) are in
computable forms. {\subsubsection{ \bf Case (II): $\nu$ is a
positive integer}}

The poles of the gammas in the integral representation of (4.35) are
as follows: $\Gamma(s) : ~ s=0,-1,-2,\cdots;~\Gamma \left(
\frac{1}{2}+s\right) :~ s =
-\frac{1}{2},-\frac{3}{2},-\frac{5}{2},\cdots ;~\Gamma (1+\nu+s ):
 ~s=-\nu-1,-\nu-2,-\nu-3,\cdots.$ \\
Some poles of $\Gamma(s)$ and $\Gamma (1+\nu+s )$ will coincide with
each other. Therefore these will be of order 2.  We note that the
poles $s=0,-1,-2,\cdots,-\nu $   are each of order 1; $s =
-\frac{1}{2},-\frac{3}{2},-\frac{5}{2},\cdots $  are each of order
1; $s=-\nu-1,-\nu-2,-\nu-3,\cdots$ are each of order 2.  Evaluating
the sum of the residues we have

\begin{thm}
If $\nu> 0$ is an integer, then for $b>0,~\alpha<1$, we have
\begin{eqnarray}
&&\int _{ 0 }^{d} x^{\nu}[1-(1-\alpha )x]^{\frac{1}{1-\alpha }}
{\rm e}^{-bx^{-\frac{1}{2}}}{\rm d}x  \nonumber\\
 &=&\frac{\Gamma \left ( \frac{1}{1-\alpha }+1 \right )}
{\sqrt{\pi} (1-\alpha )^{\nu+1 }}\bigg\{
\frac{(\pi)^{\frac{1}{2}}\Gamma(1+\nu)}{\Gamma \left
(2+\nu+\frac{1}{1-\alpha }
 \right ) }\sum_{r=0}^{\nu}\frac{\left(-1-\nu-\frac{1}{1-\alpha }
 \right )_r}{\left(\frac{1}{2}\right)_r(-\nu)_r r!}
 \left[ \frac{(1-\alpha )b^2}{4}\right]^{r} \nonumber \\
 &-&\frac{2(\pi)^{\frac{1}{2}}\Gamma\left(\frac{1}{2}+\nu\right)}{\Gamma\left
(\frac{3}{2}+\nu+\frac{1}{1-\alpha }
 \right )}\left( \frac{(1-\alpha )b^2}{4}\right)^{\frac{1}{2}}
{_1F_2}\left(-\frac{1}{2}-\nu-\frac{1}{\alpha-1};~\frac{3}{2},\frac{1}{2}-\nu;~
\frac{(1-\alpha )b^2}{4}\right) \nonumber \\
&+&\biggl(\frac{(1-\alpha)b^2}{4}\biggr)^{1+\nu}\sum_{r=0}^{\infty}
\biggl(\frac{(1-\alpha)b^2}{4}\biggr)^{r} \bigg[-\ln\left(
\frac{(1-\alpha)b^2}{4}\right)+E_r\bigg]F_r \bigg\}
\end{eqnarray}
where
\begin{equation}
E_r=\psi\left(-\frac{1}{2}-\nu-r\right)-
\psi\left(\frac{1}{1-\alpha}+1-r \right)+\psi(1+r) +\psi(2+\nu+r)
\end{equation}
and
\begin{equation}
F_r=\frac{(-1)^r(-1)^{1+\nu +r}\Gamma\left(-\frac{1}{2}-\nu
\right)\Gamma\left(-\frac{1}{1-\alpha}\right)
}{\left(\frac{3}{2}+\nu \right)_r r!(1+\nu +r)!\Gamma
\left(\frac{1}{1-\alpha}+1 \right)}
\end{equation}
\end{thm}
{\subsubsection{\bf{Case (III): $\nu$ a negative integer}} Let
$\nu=-\mu,\mu>0.$  Then proceeding as in the above cases we have the
 result
\begin{thm}
\begin{eqnarray}
&&\int _{ 0 }^{d} x^{\nu}[1-(1-\alpha )x]^{\frac{1}{1-\alpha }}
{\rm e}^{-bx^{-\frac{1}{2}}}{\rm d}x  \nonumber\\
 &=&\frac{\Gamma \left ( \frac{1}{1-\alpha }+1 \right )}
{\sqrt{\pi} (1-\alpha )^{\nu+1
}}\bigg\{\frac{-2(\pi)^{\frac{1}{2}}\Gamma\left(\frac{1}{2}+\nu\right)}{\Gamma\left
(\frac{3}{2}+\nu+\frac{1}{1-\alpha }
 \right )}\left( \frac{(1-\alpha )b^2}{4}\right)^{\frac{1}{2}}\nonumber \\
&&\times
{_1F_2}\left(-\frac{1}{2}-\nu-\frac{1}{\alpha-1};~\frac{3}{2},\frac{1}{2}-\nu;~
\frac{(1-\alpha )b^2}{4}\right) \nonumber \\
&+&\frac{\Gamma(-1-\nu)\Gamma\left(-\frac{3}{2}-\nu\right)}
{\Gamma\left(1+\frac{1}{1-\alpha}\right)}
\left[\frac{(1-\alpha)b^2}{4}\right]^{1+\nu}
\sum_{r=0}^{-\nu-2}\frac{1}{r!}
\frac{\left(-\frac{1}{1-\alpha}\right)_r}
{\left(\frac{5}{2}+\nu\right)_r(2+\nu)_r}\left[\frac{(1-\alpha)b^2}{4}\right]^r
\nonumber \\
&+&\sum_{r=0}^{\infty} \left(\frac{(1-\alpha)b^2}{4}\right)^r
\bigg[-\ln\left( \frac{(1-\alpha)b^2}{4}\right)+{E'}_r\bigg]
{F'}_r\bigg\},
\end{eqnarray}
where
\begin{equation}
{E'}_r=\psi\left(\frac{1}{2}-r\right)-
\psi\left(\frac{1}{1-\alpha}+2+\nu-r \right)+\psi(1+r) +\psi(r-\nu)
\end{equation}
and
\begin{equation}
{F'}_r=\frac{(-1)^{1+\nu }\Gamma\left(\frac{1}{2}-r \right)
 } { r!(r-\nu-1)!\Gamma\left(\frac{1}{1-\alpha}+2+\nu-r\right)}
\end{equation}
\end{thm}
 {\subsubsection{\bf{Case (IV): $\nu$ a
positive half integer}} Proceeding as before, we have
\begin{thm}
For $\nu$ a positive half integer, namely $\nu=m+\frac{1}{2},
~m=0,1,2,\cdots$
\begin{eqnarray}
&&\int _{ 0 }^{d} x^{\nu}[1-(1-\alpha )x]^{\frac{1}{1-\alpha }}
{\rm e}^{-bx^{-\frac{1}{2}}}{\rm d}x  \nonumber\\
  &=&\frac{\Gamma \left ( \frac{1}{1-\alpha }+1 \right )}
{\sqrt{\pi} (1-\alpha )^{\nu+1
}}\bigg\{\frac{(\pi)^{\frac{1}{2}}\Gamma(m+\frac{3}{2})}{\Gamma
\left (m+\frac{5}{2}+\frac{1}{1-\alpha }
 \right ) }{_1F_2}\left(-m-\frac{3}{2}-\frac{1}{1-\alpha};~\frac{1}{2},~-m-\frac{1}{2};
 ~\frac{(1-\alpha )b^2}{4}\right)\nonumber \\
 &-&\frac{2(\pi)^{\frac{1}{2}}\Gamma(1+m)}{\Gamma\left
(2+m+\frac{1}{1-\alpha }
 \right )}\left( \frac{(1-\alpha )b^2}{4}\right)^{\frac{1}{2}}
\sum_{r=0}^m \frac{(-1)^r}{r!}\frac{\left(-1-m-\frac{1}{\alpha-1
}\right)_r} {\left(\frac{3}{2}\right)_r(-m)_r}\left(\frac{(1-\alpha
)b^2}{4}\right)^r \nonumber \\
&+&\left(\frac{(1-\alpha)b^2}{4}\right)^{m+\frac{3}{2}}\sum_{r=0}^{\infty}
\left(\frac{(1-\alpha)b^2}{4}\right)^r \bigg[-\ln\left(
\frac{(1-\alpha)b^2}{4}\right)+G_r\bigg]H_r\bigg\},
\end{eqnarray}
where
\begin{equation}
G_r=\psi\left(-m-\frac{3}{2}-r\right)+
\psi\left(1+\frac{1}{1-\alpha}-r \right)+\psi(1+r) +\psi(2+m+r)
\end{equation}
and
\begin{equation}
H_r=\frac{(-1)^{1+m+r}(-1)^r\Gamma\left(-m-\frac{3}{2}
\right)\left(-m-\frac{3}{2} \right)_r} { r!(1+m
+r)!\Gamma\left(1+\frac{1}{1-\alpha}\right)
\left(1+\frac{1}{1-\alpha}\right)_r}
\end{equation}
\end{thm}
{\subsubsection{\bf{Case (V): $\nu$ a negative half integer}} Let
$\nu=-m-\frac{1}{2},m=0,1,2,\cdots.$ In this case we have
\begin{thm}
For $\nu$ a negative half integer, namely $\nu=-m-\frac{1}{2},
~m=0,1,2,\cdots$
\begin{eqnarray}
&&\int _{ 0 }^{d} x^{\nu}[1-(1-\alpha )x]^{\frac{1}{1-\alpha }}
{\rm e}^{-bx^{-\frac{1}{2}}}{\rm d}x  \nonumber\\
 &=&\frac{\Gamma \left ( \frac{1}{1-\alpha }+1 \right )}
{\sqrt{\pi} (1-\alpha )^{\nu+1
}}\bigg\{\frac{(\pi)^{\frac{1}{2}}\Gamma(\frac{1}{2}+m)}{\Gamma
\left (\frac{3}{2}-m+\frac{1}{1-\alpha }
 \right ) }{_1F_2}\left(m-\frac{1}{2}-\frac{1}{1-\alpha};~\frac{1}{2},~\frac{1}{2}+m;
 ~\frac{(1-\alpha )b^2}{4}\right)\nonumber \\
 &&+\frac{\Gamma\left(m-\frac{1}{2}\right)\Gamma(1+m)}{\Gamma\left
(1+\frac{1}{1-\alpha }
 \right )}\left( \frac{(1-\alpha )b^2}{4}\right)^{\frac{1}{2}-m}
\sum_{r=0}^{m-1} \frac{1}{r!}\frac{\left(-\frac{1}{1-\alpha
}\right)_r} {\left(\frac{3}{2}\right)_r(-m)_r}\left(\frac{(1-\alpha
)b^2}{4}\right)^r\nonumber \\
&&-2\sqrt{\pi}\sum_{r=0}^{\infty}
\left(\frac{(1-\alpha)b^2}{4}\right)^{\frac{1}{2}+r}
\bigg[-\ln\left(
\frac{(1-\alpha)b^2}{4}\right)+{G'}_r\bigg]{H'}_r\bigg\},
\end{eqnarray}
where
\begin{equation}
{G'}_r=\psi\left(-\frac{1}{2}-r\right)+
\psi\left(1+\frac{1}{1-\alpha}-m-r \right)+\psi(1+r) +\psi(1+m+r)
\end{equation}
and
\begin{equation}
{H'}_r=\frac{(-1)^m\left(-m-\frac{1}{1-\alpha} \right)_r} { r!(m
+r)!\Gamma\left(1+\frac{1}{1-\alpha}-m\right) }
\end{equation}
\end{thm}
\begin{center}
{\small {\section{\bf Behaviour of the integrals $I_{1\alpha}$ and
$I_{2\alpha}^{(d)}$ }}}
\end{center}

  The behaviour of the integral $I_{1\alpha}$ is such that
 as the value of the pathway parameter $\alpha$
changes the curve will move away from the stable
 situation ie, the strict Maxwell-Boltzmann situation(Figure 3 below).
The graphs of the integral $I_{1\alpha}$ when $\nu=1$ and at
$\alpha=1,~\alpha=1.25,~\alpha=1.35,~\alpha=1.45$ are plotted in
Figure 1. As $\alpha\geq 1.5$ the G-function  nolonger exists as it
violates the conditions. We can take other value of $\nu$ also.
\begin{center}
\resizebox{6cm}{!}{\includegraphics{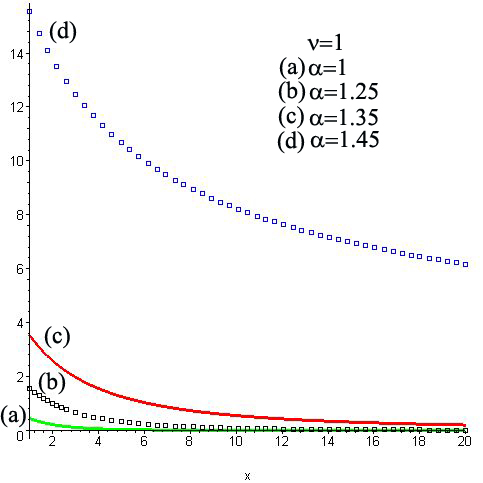}}\\
Figure 1. Behaviour of $I_{1\alpha}$ for various values of
$\alpha>1$
\end{center}
 Similarly the behaviour of the integrals $I_{2\alpha}^{(d)}$ is
such that the function moves away from the stable case and comes
closer to the origin.
\begin{center}
\resizebox{6cm}{!}{\includegraphics{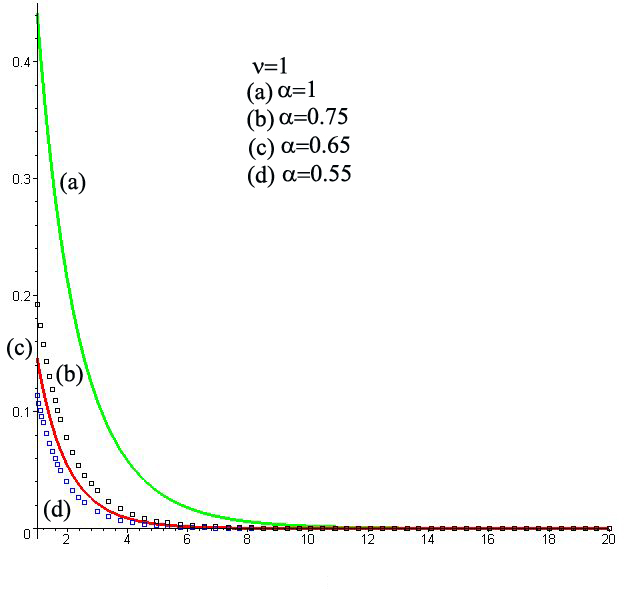}}\\
 Figure 2. Behaviour of $I_{2\alpha}^{(d)}$ for various values of $\alpha<1$
 \end{center}
As $\alpha\rightarrow1$ we get the standard situation which is done
in the series of papers of Mathai and Haubold. As
$\alpha\rightarrow1$ the two integrals will come close to the
following limiting situation.
\begin{center}
\resizebox{6cm}{!}{\includegraphics{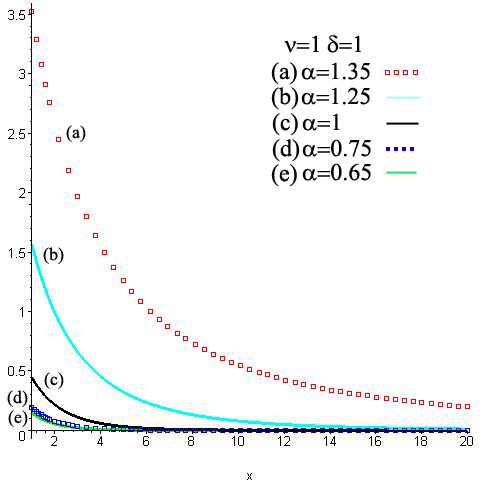}}(a)
 \resizebox{6cm}{!}{\includegraphics{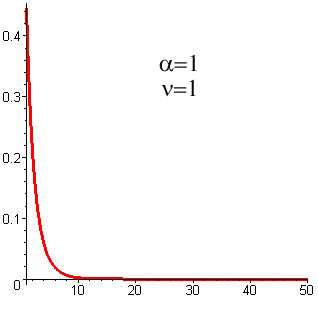}}(b)\\
  Figure 3. Maxwell-Boltzmann case or the limiting situation
$\alpha=1.$
 \end{center}
In figure 4 as the value of $\delta$ moves we can see the depletion
in the high energy tail of the Maxwell-Boltzmann situation. The
graphs of depletion in the stable situation as well as many unstable
and chaotic situations are plotted here.  The cases (i),(ii) and
(iii) show the depletion  when $\alpha=1.25,~\alpha=1.35$ and
$\alpha=1.45$ respectively and (iv) shows the depletion in the
stable situation ($\alpha=1$).
\begin{center}
\resizebox{6cm}{!}{\includegraphics{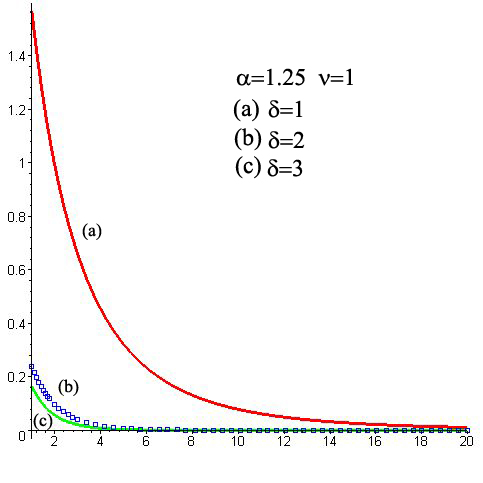}}(i)
\resizebox{6cm}{!}{\includegraphics{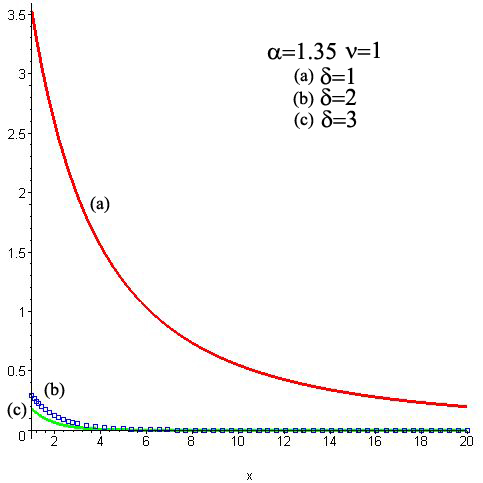}}(ii)
\resizebox{6cm}{!}{\includegraphics{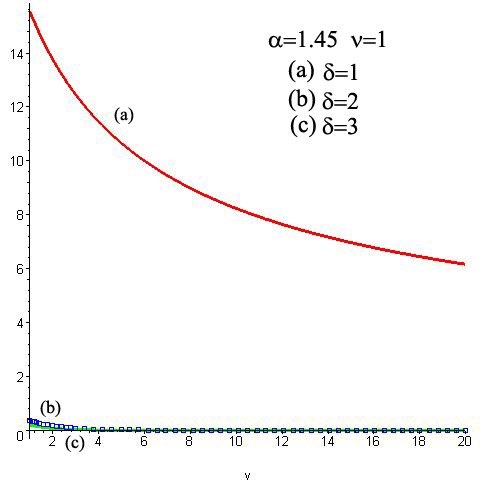}}(iii)
\resizebox{6cm}{!}{\includegraphics{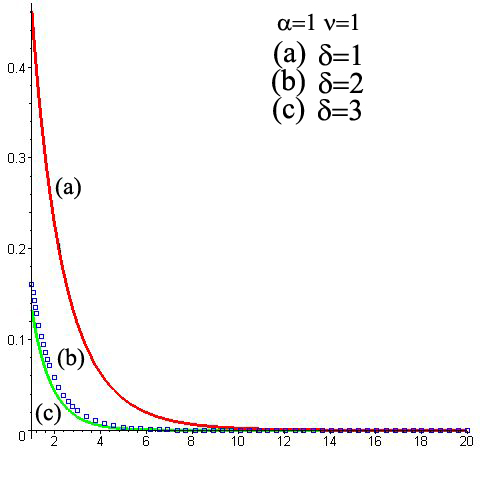}}(iv)\\
 Figure 4. Depletion for $\delta=1,2,3$ and  $\alpha=1.25,1.35,1.45,1$
\end{center}
\section{\bf{Conclusion}}

For the analytic evaluation of the probability integral for
equilibrium conditions we consider a more general form of the
reaction probability integral. We investigated in section 4 the
series representations of the extended integral, $I_{1\alpha}$ and
$I_{2\alpha}^{(d)}$ where as the pathway parameter
$\alpha\rightarrow1$ one gets the Maxwell-Boltzmann case.
$I_{1\alpha}$ for various values of $\alpha$  is plotted in Figure
1, $I_{2\alpha}^{(d)}$ for various values of $\alpha$ is plotted in
Figure 2 and in Figure 3 the limiting case,namely $\alpha=1$ case or
Maxwell-Boltzmann situation is plotted. The plotting is done by
Maple 9.
\newpage
\begin{center}{\bf Acknowledgement}\end{center}

The author would like to thank the Department of Science and
Technology, Government of India, New Delhi, for the financial assistance for this work
under project No. SR/S4/MS:287/05, and the Centre for Mathematical
Sciences for providing all facilities.
 \begin{center}{\bf References}\end{center}
{\small
\begin{description}
\item
Anderson, W.J., Haubold, H.J. and Mathai, A.M.:
 1994, Astrophysical thermonuclear functions, {\it
Astrophysics and Space Science} {\bf 214}, 49-70.
\item
Bergstroem, L., Iguri, S. and Rubinstein, H.: 1999, Constraints on
the variation of the fine structure constant from big bang
nucleosynthesis, {\it Physical Review} {\bf D60}, 045005-1-045005-9.
\item
Coraddu, M., Kaniadakis, G., Lavagno, A. Lissia, M., Mezzorani, G.,
and Quarati, P.: 1999, Thermal distributions in stellar plasmas,
nuclear reactions and solar neutrinos, {\it Brazilian Journal of
Physics} {\bf 29}, 153-168.
\item
Gell-Mann, M. and Tsallis, C. (Eds.): 2004,{\it Nonextensive
Entropy: Interdisciplinary Applications }, Oxford University Press,
New York.
\item
Haubold, H.J. and John, R.W.: 1978,   On the evaluation of
 an itegral connected with the thermonuclear reaction rate
 in closed-form, {\it Astronomische  Nachrichten} {\bf 299}, 225-232.
\item
Haubold, H.J. and John, R.W.: 1982,   On resonant thermonuclear
reaction rate integrals-closed form evaluation and approximation
considerations, {\it Astronomische  Nachrichten} {\bf 303}, 161-187.
\item
Haubold, H.J. and Kumar, D.: 2007, Extension of thermonuclear
functions through the pathway model including Maxwell-Boltzmann and
Tsallis distributions, arXiv:astro-ph/0708.2239v1
\item
Haubold, H.J. and Mathai, A.M.: 1984 , On  nuclear reaction rate theory, {\it Annalen der
Physik (Leipzig)} {\bf 41(6)}, 380-396.
\item
Haubold, H.J. and Mathai, A.M.: 1985 , The Maxwell-Boltzmannian
Approach to the Nuclear Reaction Rate Theory, {\it Fortschritte der Physik}
 {\bf 33}(11-12), 623-644.
\item
Haubold, H.J. and Mathai, A.M.: 1986, Analytic representations of
modified non-resonant thermonuclear reaction rates, {\it Journal of
Applied Mathematics and Physics (ZAMP)} {\bf 37}(5), 685-695.
\item
Haubold, H.J. and Mathai, A.M.: 1998, On thermonuclear reaction
rates, {\it Astrophysics and Space Science} {\bf 258}, 185-199.
\item
Lavagno, A. and Quarati, P.: 2002, Classical and quantum
non-extensive statistics effects in nuclear many-body problems,
Chaos, {\it Solitons and Fractals},{\bf 13}, 569-580.
\item
Lavagno, A. and Quarati, P.: 2006, Metastability of electron-nuclear
astrophysical plasmas: motivations, signals and conditions, {\it
Astrophysics and Space Science},{\bf 305}, 253-259.
\item
Kaniadakis, G., Lavagno, A. and Quarati, P: 1997, Non-extensive
statistics and solar neutrinos,  astro ph/9701118.
\item
Kaniadakis, G., Lavagno, A., Lissia, M. and Quarati, P.: 1998,
Anomalous diffusion modifies solar neutrino fluxes, {\it Physica}
{\bf A261}, 359-373.
\item
Mathai, A.M. and Haubold, H.J.:2007, Pathway model, superstatistics,
 Tsallis statistics and a generalized measure of entropy, {\it Physica}
  {\bf A375}, 110-122.
  \item
Mathai, A.M. and Haubold, H.J.:2002, Review of mathematical
techniques applicable in astrophysical reaction rate theory,, {\it
Astrophysics and Space Science}
  {\bf 282}, 265-280.
\item
Mathai, A.M. and Haubold, H.J.:1988, {\it Modern Problems in Nuclear
and Neutrino Astrophysics}, Academie-Verlag, Berlin.
\item
Mathai, A.M.: 2005, A Pathway to matrix-variate gamma and normal
densities, {\it Linear Algebra and its Applications} {\bf 396},
317-328.
\item
Mathai, A.M.: 1993, {\it A Handbook of Generalized Special Functions
for Statistical and Physical Sciences}, Clarendo Press, Oxford.
\item
Mathai, A.M. and Saxena, R.K.: 1973, {\it Generalized Hypergeometric
Functions with Applications in Statistics and Physical Sciences},
Springer-Verlag, Lecture Notes in Mathematics {\bf Vol.348}, Berlin
Heidelberg, New York.
\item
Saxena R.K., Mathai A.M. and Haubold H.J.:2004, Astrophysical
thermonuclear functions for Boltzmann- gibbs statistics and Tsallis
statistics,{\it Physica} {\bf A344},649-656.
\item
Tsallis,C.: 1988, Possible generalization of Boltzmann-Gibbs
statistics, {\it Journal of Statistical Physics}, {\bf 52}, 479-487.
\item
Tsallis, C.: 2004, What should a statistical mechanics satisfy to
reflect nature?,{\it Physica }, {\bf D193}, 3-34.
\end{description}}
\end{document}